# Scanning Microwave Microscopy of Vital Mitochondria in Respiration Buffer

Jinfeng Li[†¶], Zahra Nemati[‡¶], Kamel Haddadi[§], Douglas C. Wallace[Λ], Peter J. Burke[¶*]

[†]Department of Physics and Astronomy, [‡]Department of Chemical Engineering and Materials Science, [¶]Integrated Nanosystems Research Facility, [*]Electrical Engineering and Computer Science, University of California, Irvine, Irvine, CA 92697, USA

[§]Institute of Electronics, Microelectronics and Nanotechnology, University of Lille, Villeneuve d'Ascq, 59652 France

[Λ]Center for Mitochondrial and Epigenomic Medicine, Children's Hospital of Philadelphia and Department of Pathology and Laboratory Medicine, University of Pennsylvania, Philadelphia, PA 19104, USA.

*Abstract*—We demonstrate imaging using scanning microwave microscopy (SMM) of vital mitochondria in respiration buffer. The mitochondria are isolated from cultured HeLa cells and tethered to a solid graphene support. The mitochondria are kept vital (alive) using a respiration buffer, which provides nutrients to sustain the Krebs cycle. We verify that the mitochondria are "alive" by measuring the membrane potential using a voltage sensitive fluorescent dye (TMRE). The organelles are measured capacitively at 7 GHz. Several technical advances are demonstrated which enable this work: 1) The SMM operates in an electrophysiologically relevant liquid (hence conducting) environment; 2) The SMM operates in tapping mode, averaging the microwave reflection measurement over many tapping periods; 3) A tuned reflectometer enables increased sensitivity; 4) Variable frequencies up to 18 GHz are used; 5) In contrast with traditional matching/resonant methods that exhibit high quality factor that fail in the presence of liquids, interferometric/tuned reflectometer gives the possibility to adjust the quality factor or sensitivity even in the presence of the liquid.

## I. INTRODUCTION

The ultra-structure of mitochondria is critically related to metabolism and cell death pathways [1], [2].The inner membrane is folded and has cristae necks of diameter ~ 10 nm. The change in this ultrastructure and its relationship to cell death pathways (apoptosis) is controversial and difficult to study, as imaging with optical microscopy lacks the required spatial resolution. Electron microscopy can only provide a snapshot in time of a frozen sample. In addition to ultra-structure, mitochondria are electrically active and sustain a membrane potential of ~ 0.1 V, but there are no tools to patch clamp due to the complex nature of the ultrastructure. Real time nano-probes that have high spatial resolution and function in liquid are needed to further the field of mitochondrial biology. Atomic force microscopy in liquid alone cannot provide this information, as it only provides topological information about the surface of the organelle. The ultrastructural changes are mostly *inside* the organelle and are not imaged by AFM alone.

SMM has the potential to measure the inside of living systems, acting as a "nano-radar". In dry applications, variable frequency measurements penetrate deep into semiconductor samples (in a frequency dependent way), enabling calibrated measurements of doping profiles in all three dimensions: X and Y with the physical scan, and Z by varying the frequency. Excellent progress shows spatial resolution of 50 nm and capacitance can be calibrated to the ~100 aF scale [3]. Translating these technological advances into liquid environment presents many challenges, discussed in more depth below. Progress to date includes imaging of the outer shape of vesicles/exosomes [4] in liquid, imaging of the surfaces of cells in culture or tissue [5]–[7] or even low contrast images of the inside of CHO and E-Coli [8]. However, to date, no images of vital sub-cellular organelles, either inside or outside cells, have ever been published.

In this work, we demonstrate imaging using SMM of vital mitochondria in respiration buffer. The mitochondria are isolated from cultured HeLa cells and tethered to a solid graphene support. The mitochondria are kept vital (alive) using a respiration buffer which provide nutrients to sustain the Krebs cycle. We verify that the mitochondria are "alive" by measuring the membrane potential using a voltage sensitive fluorescent dye (TMRE) [9].

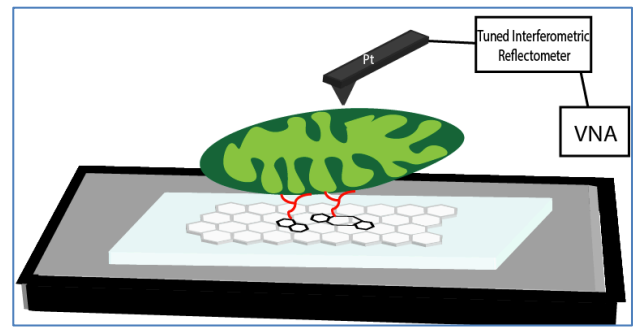

Fig. 1. Cartoon of a tethered mitochondrion onto a graphene support in a liquid environment. The live mitochondrion is then imaged via SMM tapping mode capability.

## II. MATERIALS AND METHODS

*A. Mitochondrial Isolation*

On the days of the experiment, $10^7$ HeLa cells were harvested for mitochondrial isolation. Before isolation, the confluent cells were stained with MitoTracker Green FM and TMRE for 0.5-1 hour. Mitochondria from the cultured cells were isolated using differential centrifugation steps. We followed the isolation protocol described in [10]. From this step forward through tethering and SMM imaging, the isolated mitochondria were suspended in respiration buffer (140 mM KCl, 2 mM MgCl2, 10 mM NaCl, 0.5 mM EGTA, 0.5 mM KH2PO4, 2 mM HEPES, 5 mM succinate, pH 7.2 adjusted with KOH), which results in ADP-stimulated respiration and coupled with oxygen consumption would maintain the isolated mitochondria in a vital state. The isolated mitochondrial samples were then divided into two separate aliquots, one fluorescently imaged for proof of life and the second imaged via SMM. Using an Olympus inverted microscope with two LED excitation sources (490 nm & 565 nm), we observed the red and green fluorescence signals from MitoTracker Green and TMRE. To process and analyze the images, ImageJ software was used.

*B. Graphene Device Fabrication and Functionalization*

CVD graphene was transferred onto PDMS and functionalized via a series of solution deposition methods. The graphene transfer and functionalization schemes are described in [10]. Isolated mitochondria were then loaded onto the graphene device and incubated for 15 min at 4°C, followed by a wash to remove the untethered ones before imaging.

### III. SCANNING MICROWAVE MICROSCOPE (SMM)

The scanning microwave microscope, from Keysight$^{TM}$ (model 7500) consists of an AFM interfaced with a performance vector network analyzer, as shown in Fig. 2. A microwave signal is transmitted from the PNA to a conductive AFM probe that operates in tapping mode with the sample being scanned. The probe also serves as a receiver to capture the reflected microwave signal from the contact point. By directly measuring the complex reflection coefficient, the impedance of the sample at each scanned point can be then mapped, simultaneously with the surface topography [11]. In the proposed configuration, a homemade tuned interferometric system is developed to control the interference frequency position and the level of the magnitude of the reflection coefficient. The interferometer is built up in coaxial form with a hybrid coupler, passive and variable phase-shifter and attenuator, and a low-noise amplifier.

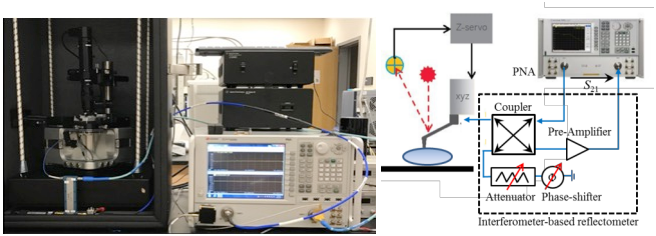

Fig. 2. The photo and diagram of the SMM setup.

According to Fig. 2, the basic principle consists of splitting the PNA source microwave signal into two coherent signals, i.e. the reference signal to the AFM tip and the interference signal.

The coupler acts as a reflectometer to separate the incident wave and the reflected wave from at the tip interface. The reflected wave is then combined with the interference signal to provide the output signal at the output of the coupler. This signal is cancelled by properly tuning the attenuator and phase-shifter. The resulting signal is amplified and measured by the PNA receiver in transmission mode. The interference signal is shown in Fig. 3. By changing the variable phase shifter and attenuator in the interferometer's reference arm, the magnitude and position of the interference peaks can be adjusted. At the best impedance match points (purple curve in the inset of Fig. 3), the system is at the best sensitivity to the tip-sample impedance. Capacitance change down to attofarad can be measured in such condition

When the tip-sample is immersed into water, there are noticeable changes we can see in Fig. 3. The $S_{21}$-spectrum shifted to the small frequency direction by ~0.2GHz. By comparing the image of standard sample at different resonance frequencies, we observed a trend of increasing in the SMM resolution as the microwave frequency going up. In addition, the resonance phenomenon gets weak when the frequency goes up. To compromise, we choose the peak at ~7GHz to image the mitochondria.

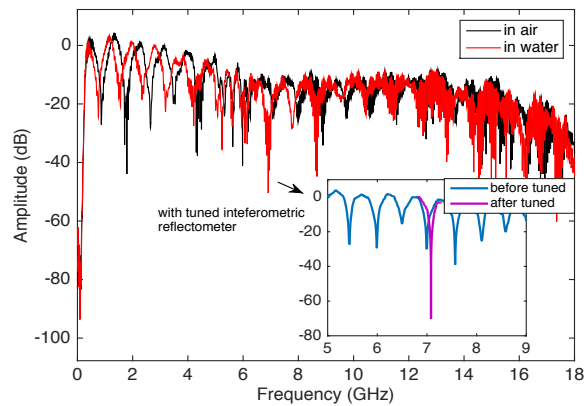

Fig. 3. The interference signals. The change of the interference signal in water vs. in air. The inset shows the tuning of the interference peak in water.

### IV. MITOCHONDRIA STUDIES

*C. Verification of Mitochondria with Optical Microscopy*

Fig. 4 shows isolated mitochondria from HeLa cancer cell lines that were fluorescently tagged. These organelles were then tethered to the surfaces of graphene coated glass substrates using a step-wise organic functionalization scheme as described in the methods section. The TMRE potentiometric dye is a cell permeant, positively charged dye that accumulates in active mitochondria due to their negative charge, and fails to do so in inactive or depolarized mitochondria given the diminished charge. As a result, a TMRE fluorescence signal is indicative of the vitality of isolated mitochondria.

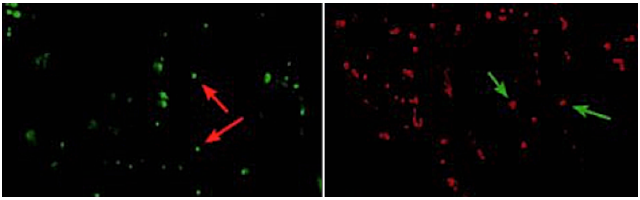

Fig. 4.  The left image shows Mito-Tracker green tagged isolated mitochondria and the right are TMRE potentiometric fluorescently tagged mitochondria. Both images have been modified with added false color.

*D. SMM characterization of Mitochondria*

The device was brought into contact with the SMM probe in tapping mode with gentle tip-sample interaction to prevent displacement or damage of mitochondria. We then did a tuning of the reflectometer and selected the inference peak at ~7 GHz (Fig. 1a) for mitochondria imaging.

We acquired an uncalibrated capacitance map of a single vital mitochondria as seen in Figure 5. The diameter of the mitochondria is ~1 µm. The graphene layer underneath has a surface roughness of a few nm, whose ripples can be seen in the topographic image. Mitochondria, as non-conductive organelles, sitting on the conductive background of graphene, give the SMM image of mitochondria a very sharp contrast. Graphene was used in anticipation of future studies that would exploit its pH sensing capability [10] simultaneously with SMM and AFM studies presented in this manuscript.

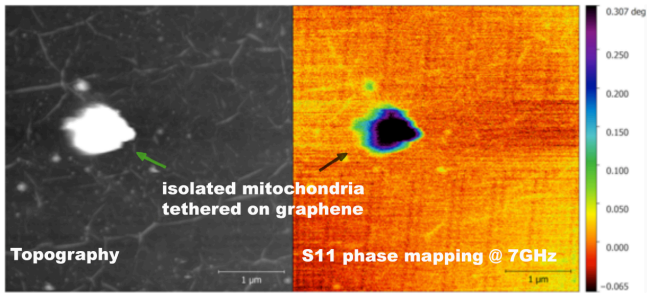

Fig. 5.  A single live mitochondrion (isolated from HeLa cell culture and tethered on graphene support) with standard topographic imaging mode (black and white image), and scanning microwave microscopy (color image)

## V. Discussion

In this work, we have presented the first SMM image of a vital sub-cellular organelle. The topographic and SMM images are consistent with the known outer morphology of mitochondria. However, the potential advantage of SMM is the ability to see inside the organelle, which has not yet been demonstrated in this work. In principle, this could provide more information than simply topographic mode. In the past, this concept has been proven with dry materials samples, able to penetrate progressively deeper into a solid as the frequency is increased [12]. It is our claim that, having demonstrated SMM imaging of a vital organelle, that this opens a new window of opportunity for future imaging of the ultra-structure of all sorts of organelles.

Although mitochondria are of particular interest in medicine and biology [13] and are highly electrically active [14] making them an excellent target for new nano-electronic imaging technologies, the technique should be applicable to other systems such as exosomes, chloroplasts, and bacteria.

## VI. Conclusion

We have demonstrated the first SMM images of vital isolated mitochondria in physiologically relevant respiration buffer. The mitochondrial capacitance is assayed and imaged. This represents proof of concept of SMM in electrophysiologically active organelles, which provides information complementary to optical and electron microscopies. Since this approach is fully functional in biological buffer, it enables studies of the changes mitochondria undergo under different chemical environments, such as cell death signals (e.g. BCL2 proteins), different metabolites, mitochondrial ROS, and many other studies, all in real time.